\documentclass{article}
\usepackage{dcase2024_techrep,amsmath,graphicx,url,times,booktabs, tabularx}
\usepackage{multirow}
\usepackage{multirow}
\usepackage{amssymb}
\usepackage{cite}
\title{FEW-SHOT BIOACOUSTIC EVENT DETECTION WITH FRAME-LEVEL EMBEDDING LEARNING SYSTEM}
\name{PengYuan Zhao$^{1}$,
      ChengWei Lu$^{1}$,
      Liang Zou$^{1}$,
      }
\address{$^1$ China University of Mining and Technology, XuZhou, China, \{pengyuan, chengwei, liangzou \}@cumt.edu.cn\\          
 }

\begin{document}

\ninept
\maketitle

\begin{sloppy}

\begin{abstract}
This technical report presents our frame-level embedding learning system for the DCASE2024 challenge for few-shot bioacoustic event detection (Task 5). We used an open-source multi-task frame-level learning system. In this work, we used log-mel and PCEN for feature extraction of the input audio, Netmamba Encoder as the information interaction network, and adopted data augmentation strategies to improve the generalizability of the trained model as well as multiple post-processing methods. Our final system achieved an F-measure score of 56.4\% in DCASE2024 task 5 Validation set. 
\end{abstract}


\section{Introduction}
\label{sec:intro}
Few-shot learning (FSL) \cite{snell2017} aims to train models that can accurately predict new data with a small amount of labeled data. Methods such as Meta-learning architectures and data augmentation techniques are commonly used to improve model generalization and increase the diversity of training data. This approach is particularly suitable for scenarios where data is scarce, costly to obtain, or difficult to label. Bioacoustic event detection (BED) field refers to the automated identification and classification of sounds produced by various organisms (such as birds, mammals, etc. ) in nature. Deep learning and transfer learning are commonly used to enhance the accuracy and generalization of sound event detection and classification. This technology plays an important role in ecological monitoring and species conservation. However, annotating bioacoustic data usually requires significant time and effort from experts in the field. Few-shot bioacoustic event detection (FSBED) \cite{nolasco2023learning, liang2024mind} combines the advantages of few-shot learning, requiring only a small amount of supervised data to accomplish bioacoustic event detection, thereby saving significant time and labor costs. This technology has important applications in studying animal populations and their behaviors. 

To improve model efficiency, we use the NetMamba Encoder\cite{gu2023mamba} as the Information exchange network instead of the Transformer encoder. Mamba is a linear-time state space model for sequence modeling that has achieved significant success in various fields, including natural language processing\cite{he2024densemamba} and computer vision\cite{zhu2024vision}. Through innovative design, the Mamba network addresses many issues encountered by RNNs, LSTMs, and Transformers \cite{schafer2024animal2vec, zou2024multitask} when processing long sequential signals, such as high computational complexity, large memory consumption, and gradient vanishing. It excels in capturing long-term dependencies, computational efficiency, and memory efficiency, providing a more efficient and stable solution for long sequential signal processing tasks. This indicates that applying Mamba to the field of few-shot bioacoustic event detection has great potential. To achieve superior performance, inference efficiency, and few-shot learning capability, we chose NetMamba\cite{wang2024netmamba} among the various Mamba variants.

\section{METHODOLOGY}
\label{sec:format}

The multi-task frame-level detection training model we use is shown in Figure \ref{fig:Training Framework}. First, the input audio clip is processed and feature extraction is performed. The features are then input into the embedding extraction network. The frame-level embedding representation is fused and then used for sound event detection (SED) and foreground/background classification tasks. Finally, the embedding of the query set is multiplied by W and the prediction result is obtained through the softmax function. In the evaluation set, our method1 system base on logmel feature, which achieves an F1 score of 44.07\%. The method2 system base on PCEN feature, which achieves an F1 score of 56.4\%. Both systems perform well on validation sets such as HB and ME, but perform poorly on validation sets such as RD and PW.

\begin{figure}[h]
  \centering
  \centerline{\includegraphics[width=7cm]{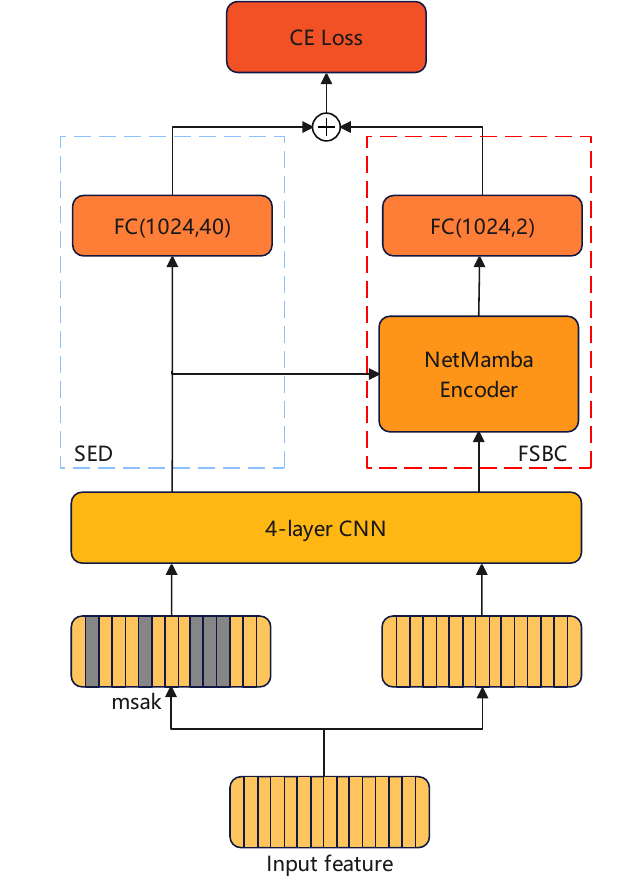}}
  \caption{Training Framework}
  \label{fig:Training Framework}
\end{figure}

\subsection{Multi-task Training Strategy}
The training framework, as depicted in Figure \ref{fig:Training Framework}, employs a multi-task frame-level model. Initially, the input audio segments are processed to extract Log-mel features. These features, along with the masked Log-mel features, are fed into an embedding extraction network consisting of four CNN Blocks (a sequence of Convolutional Layer, Batch Normalization, and ReLU activation). In the Sound Event Detection (SED) branch, the generated embedding features are utilized. In the Foreground/ Background Classification (FBC) branch, the features from the SED branch are fused with those from the FBC branch and subsequently input into the NetMamba Encoder. Ultimately, we adopt Cross-Entropy (CE) loss in both the SED and FBC branches, denoted as L1 and L2 respectively. These losses are computed using the following formulas:

\begin{equation}
  \label{l1}
    l_1 = -\frac{1}{N} \sum_{i=1}^{n} \sum_{j=1}^{c} ((Y_i = j) \odot M_i) \log(f_\theta(X_i))
\end{equation}
\begin{equation}
  \label{l2}
    l_2 = \sum_{i=1}^{n} \sum_{j=0}^{1} ((A_i = j) \odot M_i) \log(f_\theta(X_i, X_t)) \log(f_\theta(X_i))
\end{equation}
\begin{equation}
  \label{l3}
    l_{\text{total}} = l_1 + l_2
\end{equation}

\textbf{State Space Models} As the key components of Mamba, State Space Models (SSMs) represent a contemporary category of sequence models within deep learning that share broad connections with Recurrent Neural Networks (RNNs) and Convolutional Neural Networks (CNNs). Drawing inspiration from continuous systems, SSMs are commonly structured as linear Ordinary Differential Equations (ODEs) which establish a mapping from an input sequence $x\left ( t\right )\in {\mathbb{R}}^{N}$ to an output sequence $y\left ( t\right )\in {\mathbb{R}}^{N}$ via an intermediate latent state $h\left ( t\right )\in {\mathbb{R}}^{N}$:
\begin{equation}
  \label{l4}
    \begin{matrix}{h}^{\prime}\left ( t\right )=\mathbf{A}h\left ( t\right )+\mathbf{B}x\left ( t\right )
\\ 
y\left ( t\right )=\mathbf{C}h\left ( t\right )
\end{matrix}
\end{equation}
where $\mathbf{A}\in {\mathbb{R}}^{N\times N}$ represents the evolution parameter, while $\mathbf{B}\in {\mathbb{R}}^{N\times 1}$ and $\mathbf{C}\in {\mathbb{R}}^{1\times N}$ are the projection parameters.

\textbf{NetMamba Encoder} The NetMamba decoder use unidirectional Mamba blocks, as illustrated in Figure \ref{fig:NetMamba Enconder}. For a given input token sequence $ X_{t-1}$ with a batch size of $B$ and sequence length $L$ from the ($t-1$)-th NetMamba block, the decoder first normalize it and then project it linearly into $x$ and $z$, both with dimension size of $E$. It subsequently apply causal 1-$D$ convolution to $x$, resulting in ${x}^{\prime}$. Based on ${x}^{\prime}$, the decoder compute the input-dependent step size $\Delta$, as well as the projection parameters $B$ and $C$ having a dimension size of $N$. It then discretize $A$ and $B$ using $\Delta$. Following this, it calculate $y$ employing a hardware-aware SSM. Finally, $y$ is gated by $z$ and added residually to $X_{t-1}$, resulting in the output token sequence $X_{t}$ for the $t$-th NetMamba block. 

\subsection{Multi-task Fine-tuning Strategy}
After investigating a large number of existing results, we found that the fine-tuning strategy is helpful in improving the detection accuracy and generalization of the model. So we also fine-tune the SED branch and the SFBC
branch respectively on the reconstructed supports.

\textbf{SED Branch} Initially, we define a binary classifer based on the reconstructed supports set (${Support}_{1}$ and ${Support}_{2}$). Subsequently, we employ the pseudo-labels generated in the initial stage to further refine the model. This two-step process is iterated for a predetermined number of cycles.

\textbf{SFBC Branch} We select the windows in ${Support}_{1}$ containing multiple different $POS$ as TC-Vector and feed them into the lightweight feature extractor. We obtain the $POS$ center by meaning the masked POS frames and utilize the NetMamba Encoder Layer transferred from the training stage to build a self-attention with the $POS$ in ${Support}_{2}$. 

\begin{figure}[t]
  \centering
  \centerline{\includegraphics[width=5cm]{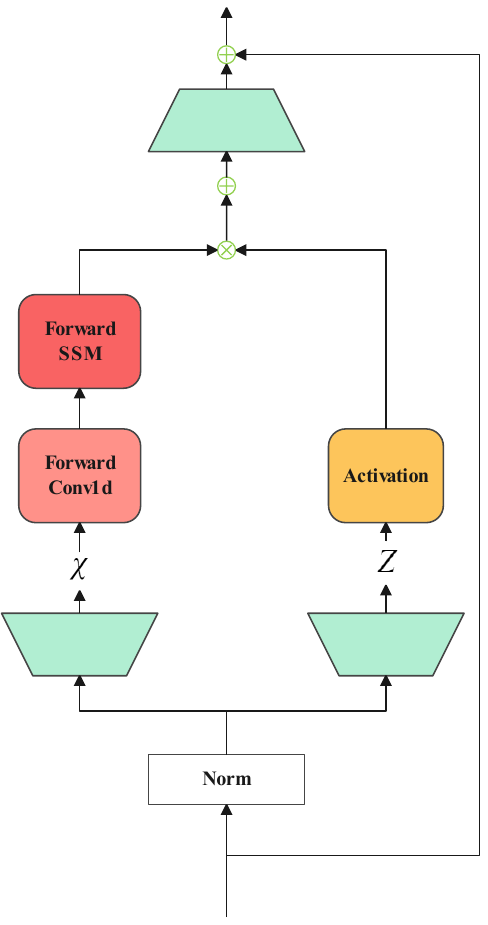}}
  \caption{NetMamba Encoder}
  \label{fig:NetMamba Enconder}
\end{figure}

\begin{table*}[htbp] 
	\centering
	\caption{Dcase Task 5 Results}
	\begin{tabular}{c c c c|c c c}
	\hline
	\multirow{2}{*}{\textbf{$System$}} & 
        \multicolumn{3}{c|}{\textbf{$Dcase 2023 Task 5$}} & \multicolumn{3}{c}{\textbf{$Dcase 2024 Task 5$}} \\
	   & \textbf{$Precision(\%)$} & \textbf{$Recall(\%)$} & 
         \textbf{$F1-score(\%)$} & \textbf{$Precision(\%)$} & \textbf{$Recall(\%)$} & \textbf{$F1-score(\%)$} \\
	   \hline
		Baseline & 36.34 & 24.96 & 29.59 & 56.18 & 48.64 & 52.14 \\
		Method1 & 76 & 67.24 & 71.35 & 35.85 & 57.19 & 44.07 \\
		Method2 & 75.68 & 69.39 & 72.4 & 61.16 & 52.32 & 56.4 \\
		\hline
	 \end{tabular} 
        \label{table1}
\end{table*}

\section{EXPERIMENTS}
\label{sec:pagelimit}

\subsection{Data Preparation}
We use the official training dataset, and we use an FFT size of 1024, hop{\_}len of 256, a Mel number of 128, and a sampling rate of 22.05kHz to calculate the Logmel spectrum features. A sliding window segmentation of Logmel is performed using a window length of 431 and a window shift of 86, and each frame in the window is labeled according to the annotation csv.

\subsection{Data Augmentation}
We observed that the quality and quantity of training data have a significant impact on the performance of the model, so we screened the training set. We observed that in the Western Mediterranean Birds (WAW) data set, there are many types of calls and long duration, but we observed that there are some labels are incorrectly labeled, so we first train the model on the training data that does not contain WAW, and then use the trained model to predict the data with weak WAW annotation information to enhance the training data.

During the training phase, we added GussionNoise to the Logmel feature. At the same time, for the time domain and frequency domain of the feature, we used the FrequencyMasking object to randomly mask a certain number of frequency channels in the input data, and used the TimeMasking object to randomly mask the time frames in the input data to achieve data enhancement and improve the model's generalization ability for different audio features.


\subsection{Post Processing}
We extract the predicted probability, minimum detection length (MFL), threshold and average event length of the model evaluation results. In particular, for the adjustment of the threshold, a strategy is set: subtract 0.05 from the threshold of the model output. If the result is lower than 0.5, the threshold is set to 0.5. This can avoid false detections caused by too low threshold. The predicted probabilities are then binarized, setting the probability above the adjusted threshold as 1 and the rest as 0. A convolution operation is used to identify significant change points in the binarized probabilities, which represent the beginning and end of events. In order to further optimize event detection, we introduced the non-maximum suppression (NMS) \cite{neubeck2006efficient} algorithm and set an IoU threshold of 0.7 to remove events that overlap or are close to each other and retain the best event detection results.

For frequently occurring short events, we have added a step of processing. If the average event length is greater than 2 times the minimum frame length (for example, 2 * 87 frames), and the interval between consecutive events is less than 87 frames, and the average predicted probability is greater than 0.5, we reset the predicted probabilities of these consecutive events to 1 to enhance the model's ability to detect such events. Finally, we implemented a temporal smoothing technique using a moving average with a window size of 5 frames to smooth the prediction probabilities and reduce noise. Additionally, we used median filtering to further reduce outliers in the predictions.
\bibliographystyle{IEEEtran}
\bibliography{refs}

\begin{thebibliography}{10}
\providecommand{\url}[1]{#1}
\def\UrlFont{\rmfamily}
\providecommand{\newblock}{\relax}
\providecommand{\bibinfo}[2]{#2}
\providecommand\BIBentrySTDinterwordspacing{\spaceskip=0pt\relax}
\providecommand\BIBentryALTinterwordstretchfactor{4}
\providecommand\BIBentryALTinterwordspacing{\spaceskip=\fontdimen2\font plus
\BIBentryALTinterwordstretchfactor\fontdimen3\font minus \fontdimen4\font\relax}
\providecommand\BIBforeignlanguage[2]{{%
\expandafter\ifx\csname l@#1\endcsname\relax
\typeout{** WARNING: IEEEtran.bst: No hyphenation pattern has been}%
\typeout{** loaded for the language `#1'. Using the pattern for}%
\typeout{** the default language instead.}%
\else
\language=\csname l@#1\endcsname
\fi
#2}}

\bibitem{snell2017}
J.~Snell, K.~Swersky, and R.~Zemel, ``Prototypical networks for few-shot learning,'' \emph{Advances in neural information processing systems}, vol.~30, 2017.

\bibitem{nolasco2023learning}
I.~Nolasco, S.~Singh, V.~Morfi, V.~Lostanlen, A.~Strandburg-Peshkin, E.~Vida{\~n}a-Vila, L.~Gill, H.~Pamu{\l}a, H.~Whitehead, I.~Kiskin, \emph{et~al.}, ``Learning to detect an animal sound from five examples,'' \emph{Ecological informatics}, vol.~77, p. 102258, 2023.

\bibitem{liang2024mind}
J.~Liang, I.~Nolasco, B.~Ghani, H.~Phan, E.~Benetos, and D.~Stowell, ``Mind the domain gap: a systematic analysis on bioacoustic sound event detection,'' \emph{arXiv preprint arXiv:2403.18638}, 2024.

\bibitem{gu2023mamba}
A.~Gu and T.~Dao, ``Mamba: Linear-time sequence modeling with selective state spaces,'' \emph{arXiv preprint arXiv:2312.00752}, 2023.

\bibitem{he2024densemamba}
W.~He, K.~Han, Y.~Tang, C.~Wang, Y.~Yang, T.~Guo, and Y.~Wang, ``Densemamba: State space models with dense hidden connection for efficient large language models,'' \emph{arXiv preprint arXiv:2403.00818}, 2024.

\bibitem{zhu2024vision}
L.~Zhu, B.~Liao, Q.~Zhang, X.~Wang, W.~Liu, and X.~Wang, ``Vision mamba: Efficient visual representation learning with bidirectional state space model,'' \emph{arXiv preprint arXiv:2401.09417}, 2024.

\bibitem{schafer2024animal2vec}
J.~Sch{\"a}fer-Zimmermann, V.~Demartsev, B.~Averly, K.~Dhanjal-Adams, M.~Duteil, G.~Gall, M.~Fai{\ss}, L.~Johnson-Ulrich, D.~Stowell, M.~Manser, \emph{et~al.}, ``animal2vec and meerkat: A self-supervised transformer for rare-event raw audio input and a large-scale reference dataset for bioacoustics,'' \emph{arXiv preprint arXiv:2406.01253}, 2024.

\bibitem{zou2024multitask}
Z.~Liang, Y.~Genwei, W.~Ruoyu, D.~Jun, L.~Meng, G.~Tian, and F.~Xin, ``Multitask frame-level learning for few-shot sound event detection,'' 2024.

\bibitem{wang2024netmamba}
T.~Wang, X.~Xie, W.~Wang, C.~Wang, Y.~Zhao, and Y.~Cui, ``Netmamba: Efficient network traffic classification via pre-training unidirectional mamba,'' \emph{arXiv preprint arXiv:2405.11449}, 2024.

\bibitem{neubeck2006efficient}
A.~Neubeck and L.~V. Gool, ``Efficient non-maximum suppression,'' in \emph{18th international conference on pattern recognition (ICPR'06)}.\hskip 1em plus 0.5em minus 0.4em\relax IEEE, 2006.

\end{thebibliography}

\end{sloppy}
\end{document}